\documentclass[12pt]{article}
\usepackage{eurosym}
\usepackage{graphicx}
\usepackage[utf8]{inputenc}
\usepackage[T1]{fontenc}
\usepackage{indentfirst}
\usepackage[margin=0.8in,nomarginpar]{geometry}
\usepackage[final]{hyperref}
\usepackage{amsmath}
\usepackage{hyperref}
\usepackage{cite}
\usepackage{subcaption}
\usepackage{caption}
\usepackage{amssymb}
\usepackage{multirow}
\usepackage{color}

\hypersetup{
colorlinks=true,
linkcolor=blue,
citecolor=blue,
filecolor=magenta,
urlcolor=blue
}

\begin{document}

\title{On Dunkl-Bose-Einstein Condensation in Harmonic Traps}
\author{A. Hocine\thanks{%
a.hocine@univ-chlef.dz} \\
Laboratory for Theoretical Physics and Material Physics Faculty of Exact \\
Sciences and Informatics, Hassiba Benbouali University of Chlef, Algeria.
\and B. Hamil\thanks{%
hamilbilel@gmail.com} \\
D\'{e}partement de physique, Facult\'{e} des Sciences Exactes, \\
Universit\'{e} Constantine 1,Constantine, Algeria. \and F. Merabtine\thanks{%
f.merabtine@univ-chlef.dz} \\
Laboratory for Theoretical Physics and Material Physics Faculty of Exact \\
Sciences and Informatics, Hassiba Benbouali University of Chlef, Algeria.
\and B. C. L\"{u}tf\"{u}o\u{g}lu\thanks{%
bekir.lutfuoglu@uhk.cz } \\
Department of Physics, University of Hradec Kr\'{a}lov\'{e},\\
Rokitansk\'{e}ho 62, 500 03 Hradec Kr\'{a}lov\'{e}, Czechia. \and M. Benarous%
\thanks{%
m.benarous@univ-chlef.dz} \\
Laboratory for Theoretical Physics and Material Physics Faculty of Exact \\
Sciences and Informatics, Hassiba Benbouali University of Chlef, Algeria. }
\date{}
\maketitle

\begin{abstract}
The use of the Dunkl derivative, which is defined by a combination of the
difference-differential and reflection operator, allows the classification
of the solutions according to even and odd solutions. Recently, we
considered the Dunkl formalism to investigate the Bose-Einstein condensation
of an ideal Bose gas confined in a gravitational field. In this work, we
address a similar problem and examine an ideal Bose gas trapped by a
three-dimensional harmonic oscillator within the Dunkl formalism. To this
end, we derive an analytic expression for the critical temperature of the N
particle system, discuss its value at large-N limit and finally derive and
compare the ground state population with the usual case result. In addition,
we explore two thermal quantities, namely the Dunkl-internal energy and the
Dunkl-heat capacity functions. The Wigner parameter of the Dunkl formalism
can be successfully used to obtain a better agreement between experimental
and theoretical results.
\end{abstract}

\section{Introduction}

The influence of external potential on the Bose-Einstein condensate (BEC),
has been extensively discussed in the literature. For instance, in \cite%
{Bagnato}, Bagnato et al calculated a three-dimensional ideal Boson gas
system's critical temperature and ground state population by using a generic
power-law potential energy. In \cite{gersch}, Gersch discussed how the
gravitational field affects the Bose-Einstein gas system's thermodynamics.
In \cite{widom}, Widom provided theoretical evidence of the BEC for an ideal
Bose liquid trapped by a gravitational field. Alike, in \cite{baranov},
Baranov et al investigated the influence of the gravitational field on the
two-dimensional BEC of atoms that are confined in a rectangular well. In 
\cite{rivas}, Rivas et al determined the BEC temperature for two distinct
trapping scenarios and discussed how the transition temperature is modified
when considering a homogeneous gravitational field. In \cite{lie}, Liu et al
examined a one-dimensional non-interacting Bose gas system in the presence
of a uniform gravitational field, and they derived the BEC temperature and
the condensate fraction by using the semiclassical approach. Subsequently in 
\cite{du}, Du et al handled the same problem in two and three dimensions,
and they obtained new features beyond the results of Liu et al. Harmonic
potential traps were also subjected as an external potential. In \cite%
{kirsten 1}, Kirsten et al discussed the BEC of atomic gases in a spin-0
system with harmonic oscillator potential energy. In \cite{katrel}, Ketterle
et al examined the BEC for one and three-dimensional nonrelativistic systems
under isotropic harmonic oscillator potential, while in \cite{mullin} Mullin
handled the same problem in two dimensions. Later, in \cite{mullinS} he
discussed the problem in a more generalized form. In outstanding works \cite%
{zeng 1, zeng 2}, Qi-Jun Zeng et al studied the BEC of a two and
three-dimensional harmonically trapped system in the context of the
q-deformed bosons theory.

A question that would historically be assumed to form the basis of Dunkl
derivation and formalism was posed by Eug\`ene Wigner in the middle of the
last century: "Can the dynamics of a quantum mechanical system produce
canonical commutation relations? " \cite{Wigner}. Although Wigner's
conclusion was negative, because of an extra constant parameter that forbids
a unique solution, one year later L. M. Yang managed to present a unique
solution by considering the one-dimensional quantum harmonic oscillator with
several restrict conditions \cite{Yang}. According to Yang, if one
introduces a reflection operator into the conventional Heisenberg algebra,
in other words, if one deforms the conventional Heisenberg algebra by 
\begin{align}
\left[\hat{x},\hat{p}\right]=i\hbar\left(1+\theta\hat{R}\right),
\end{align}
then, a unique solution is always achieved. Here, $\theta$ is the Wigner
parameter, $\hat{p}$ is the usual quantum mechanical momentum operator, and $%
\hat{R}$ is the reflection operator satisfying the following properties \cite%
{chung} 
\begin{equation}
\hat{R}_{i}f\left( x_{j}\right) =\delta _{ij}f\left(
-x_{j}\right),\quad\quad \hat{R}_{i}\frac{d}{dx_{j}}=-\delta _{ij}\hat{R}_{i}%
\frac{d}{dx_{j}}, \quad\quad \hat{R}_{i}\hat{R}_{j}=\hat{R}_{j}\hat{R}_{i}.
\end{equation}
It is worthwhile mentioning that this representation is not unique in the
position space. A particularly interesting representation can be found using
the Dunkl operator, $\hat{D}$, \cite{Dunkl1} 
\begin{equation}
\hat{D}_j\equiv \frac{i}{\hbar }\hat{p}=\frac{d}{dx_{j}}+\frac{\theta _{j}}{%
x_{j}}(1-\hat{R}_{j}) ,\text{ \ \ }j=1,2,3.
\end{equation}
which is a combination of differential and difference operators \cite{mota}.
This representation found many applications in various mathematical \cite%
{hachkm, Dunkl2, bie} as well as physical \cite{kakei, Lapointe, G1, G2, G3,
G4, Mik1, Mik2, Mik3} problems. Recently, there has been a growing interest
in using the Dunkl operator in the investigation of quantum mechanical
problems in both relativistic and non-relativistic regimes \cite%
{Sargolzaeipor, Ghaz, Mota1, Chung1, Chungrev, ubri, Kim , Hassan, Dong,
Mota2, Mota3, Merad, Bilel1, Bilel2, Bilel5, HHS, Bilel3, Merabtine}. The
reflection operator embodied in the Dunkl operator allows authors to
classify the solutions of the Dunkl-Schr\"odinger \cite{Chung1, HHS, Mota3},
the Dunkl-Dirac \cite{Sargolzaeipor, Ghaz, Mota1, Bilel5 }, the
Dunkl-Klein-Gordon \cite{Mota2, Merad, Bilel1, Bilel2}, and the
Dunkl-Duffin-Kemmer-Petiau \cite{Merad} equations by parity.

This year we studied the ideal Bose gas condensation using the Dunkl
formalism in two stages, taking into account the presence and absence of a
gravitational field \cite{Bilel3, Merabtine}. In the present manuscript, we
intend to extend these works by examining three-dimensional harmonically
trapped ideal Bose gas. We believe the free parameter of the Dunkl formalism
can be used to construct a better fit between the experimental and
theoretical results. We construct the manuscript as follows: In section 2,
we derive the BEC temperature and ground state population number in the
Dunkl formalism. In section 3, we obtain the Dunkl-internal energy and Dunkl
heat capacity functions of the system. In the last section, we conclude the
manuscript.

\section{Ideal Bose gas trapped in harmonic oscillator potential and Dunkl
formalism}

Let us consider a Bose gas composed of $N$ neutral atoms which are trapped
by a three-dimensional harmonic potential of the form 
\begin{equation}
V\left( x,y,z\right) =\frac{m\omega _{1}^{2}}{2}x^{2}+\frac{m\omega _{2}^{2}%
}{2}y^{2}+\frac{m\omega _{3}^{2}}{2}z^{2},
\end{equation}%
where $m$ and $\omega _{i}$ correspond to the mass of the atoms and their
trap frequencies. In this case, the total energy can be given by the sum of
the single-particle energies \cite{gro1} 
\begin{equation}
E_{n_{1},n_{2},n_{3}}=\hbar \left( \omega _{1}n_{1}+\omega _{2}n_{2}+\omega
_{3}n_{3}\right) +E_{0},
\end{equation}%
where $n_{i}=0,1,2,...$, $i=0,1,2...$ . Here, the zero-point energy is 
\begin{equation}
E_{0}=\frac{\hbar }{2}\left( \omega _{1}+\omega _{2}+\omega _{3}\right) .
\end{equation}%
In the Dunkl formalism, the number of condensed (ground state) and thermal
(excited states) particles are given the grand canonical ensemble as follows 
\cite{Merabtine}: 
\begin{eqnarray}
N_{0}^{D} &=&\frac{2}{z^{-2}-1}+\frac{(1+2\theta )}{z^{-(1+2\theta )}+1},
\label{N0} \\
N_{e}^{D} &=&\sum_{i\neq 0}\left( \frac{2}{e^{2\beta E_{i}}z^{-2}-1}+\frac{%
(1+2\theta )}{e^{\beta (1+2\theta )E_{i}}z^{-(1+2\theta )}+1}\right) .
\label{Ne}
\end{eqnarray}%
Here, $\beta =(k_{B}T)^{-1}$; $k_{B}$ is the Boltzmann's constant, and $%
z=e^{\beta \left( \mu -E_{0}\right) }$ is the fugacity of the system. It is
worth noting in three dimensions, the most general form of the Dunkl
formalism should be given with three different Wigner parameters, however,
for simplicity we assume that they are the same and we denote them $\theta $%
. Also, we shift the ground state energy to zero, with the replacement $\mu
-E_{0}\rightarrow \mu $, for simplifying the formulae.

\begin{figure}[hbtp]
\centering
\includegraphics[scale=0.7]{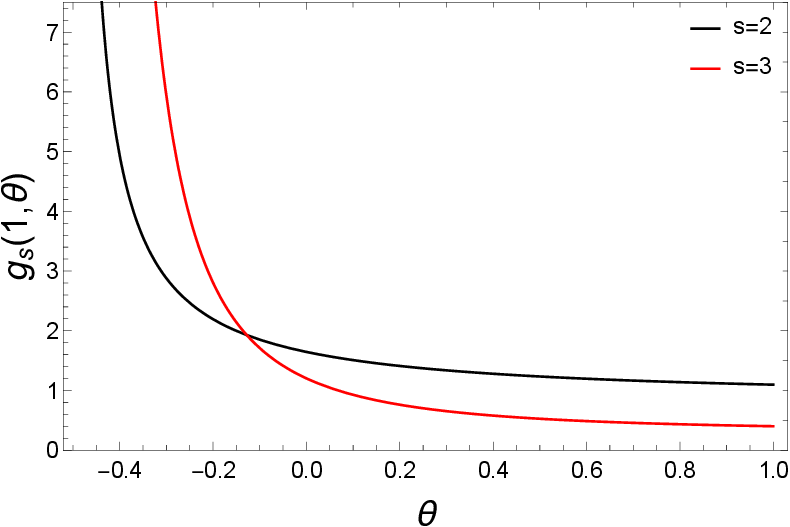}
\caption{The Dunkl-Bose function versus the Wigner parameter.}
\label{fig1}
\end{figure}

The analytical evaluation of the given sum in Eq. \eqref{Ne} is quite
difficult. To circumvent this issue, one may substitute the discrete sum
with a weighted integral, $\sum\to\int \rho(E)dE$. Here, $\rho (E)$ is the
density of states given by 
\begin{equation}
\rho \left( E\right) =\frac{1}{2}\frac{E^{2}}{\left( \hbar \Omega \right)^{3}%
}+\gamma \frac{E}{\left( \hbar \Omega \right) ^{2}},  \label{R}
\end{equation}%
where $\Omega =\left( \omega _{1}\omega _{2}\omega _{3}\right) ^{1/3}$ is
the average frequency value of the harmonic trap, and $\gamma$ is a
coefficient depends on the values of the individual frequencies of the
harmonic trap $\left( \omega _{1},\omega _{2},\omega _{3}\right)$ to be
determined numerically. In \cite{gro1}, authors calculated the $\gamma$
factor for an isotropic oscillator and they found it to be equal to $3/2$.
It should be emphasized that the above substitution is valid as long as the
number of particles is large and the spacing between energy levels is small
enough. Following some simple algebra, we obtain the total number of
particles 
\begin{equation}
N=N_{0}^{D}+2\int_{0}^{\infty }\frac{\rho \left( E\right) dE}{e^{2\beta
E}z^{-2}-1}+(1+2\theta )\int_{0}^{\infty }\frac{\rho \left( E\right) dE}{%
e^{\beta (1+2\theta )E}z^{-(1+2\theta )}+1} ,  \label{uuu}
\end{equation}
which, after integration, becomes 
\begin{align}
N& =N_{0}^{D}+\frac{1}{4}\left( \frac{k_B T}{\hbar \Omega }\right)
^{3}\left\{ g_{3}(z^{2})-\frac{4}{\left( 1+2\theta \right) ^{2}}g_{3}\left(
-z^{1+2\theta }\right) \right\} +  \notag \\
& \frac{\gamma }{2}\left( \frac{k_B T}{\hbar \Omega }\right) ^{2}\left\{
g_{2}(z^{2})-\frac{2}{\left( 1+2\theta \right) }g_{2}\left( -z^{1+2\theta
}\right) \right\} .  \label{Is}
\end{align}%
Here, the function $g_{s}(z)$ is the Bose (Polylogarithmic) function defined
by 
\begin{equation}
g_{s}(z)=\frac{1}{\Gamma (s)}\int_{0}^{\infty }\frac{x^{s-1}}{e^{x}z^{-1}-1}%
dx.
\end{equation}%
By employing the property,%
\begin{equation}
g_{s}(z)+g_{s}(-z)=2^{1-s}g_{s}(z^{2}),
\end{equation}
we restate Eq. \eqref{Is} 
\begin{equation}
N=N_{0}^{D}+\left( \frac{k_BT}{\hbar \Omega }\right) ^{3}g_{3}\left(
z,\theta \right) +\gamma \left( \frac{k_B T}{\hbar \Omega }\right)
^{2}g_{2}\left( z,\theta \right) ,  \label{s}
\end{equation}
with a new function 
\begin{equation}
g_{s}\left( z,\theta \right) =g_{s}(z)+g_{s}(-z)-\frac{1}{\left( 1+2\theta
\right) ^{s-1}}g_{s}\left( -z^{1+2\theta }\right) ,  \label{GS}
\end{equation}%
which may be called the Dunkl-Bose function. In the limit of $\theta\to 0$,
the Dunkl-Bose function reduces to the usual Bose function, so that Eq. %
\eqref{s} becomes the same as the Eq. (4) of reference \cite{gro1}. Before
the investigation of the Dunkl-BEC temperature, we would like to demonstrate
the properties of the Dunkl-Bose function. To this end, we plot the
Dunkl-Bose function for $s=2$ and $s=3$ versus the Wigner parameter in Fig. %
\ref{fig1}. 

\begin{figure}[hbtp]
\centering
\includegraphics[scale=0.7]{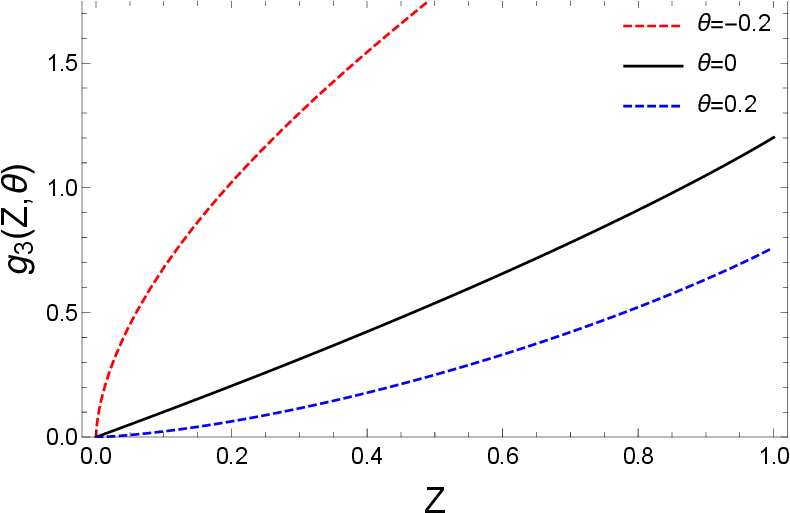}
\caption{The Dunkl-Bose function, $g_{3}(z,\protect\theta )$, versus $z$ for
different Wigner parameters.}
\label{fig2}
\end{figure}

\newpage We observe that the Dunkl-Bose function decreases as the Wigner
parameter increases. Then, we depict the Dunkl-Bose function of $s=3$ for
three different Wigner parameters versus $z$ in Fig. \ref{fig2}. 

We observe that the Dunkl-Bose function is a monotonically increasing
function. We see that for a positive value of the Wigner parameter, the
Dunkl-Bose function takes smaller values compared to the standard Bose
function. For $\theta <0$, this behavior changes oppositely, and the
Bose-Dunkl function becomes greater than the usual Bose function. These
properties are the characteristic behavior of the functions, and thus,
independent of the order $s$.

\begin{figure}[htbp]
\centering
\includegraphics[scale=0.7]{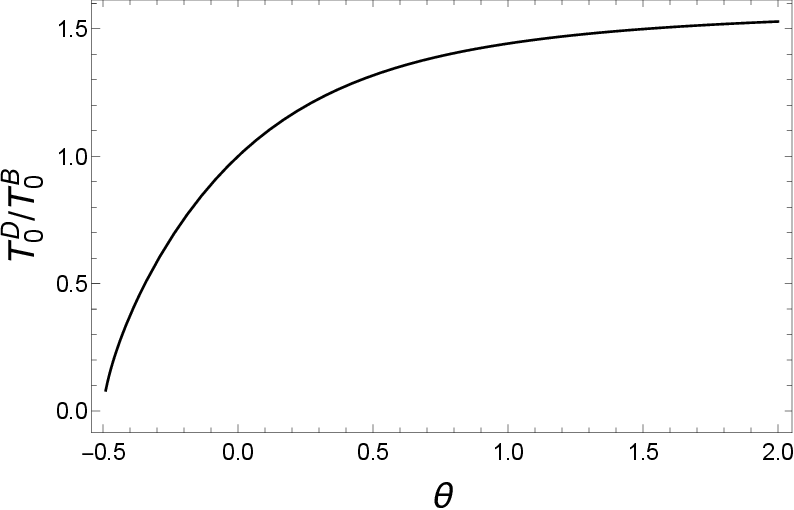}
\caption{The variation of $\frac{T_{0}^{D}}{T_{0}^{B}}$ versus $\protect%
\theta $.}
\label{fig3}
\end{figure}

Now, let us focus on the condensation temperature. We know that when the
temperature decreases to the condensation temperature, $T_{c}$, the
particles will condense in the ground state of the trap. In such a case of
the condensation onset, the system has the state of $N_{0}^D \simeq 0$ and $%
z \simeq 1$. Therefore, we can express the Dunkl-BEC temperature, $T_{c}^{D}$%
, as 
\begin{eqnarray}
T_c^D \simeq \frac{\hbar \Omega}{k_B} \bigg[\frac{N}{g_3(1,\theta)}\bigg]^
{1/3} \Bigg\{1-\frac{\gamma}{3} \Bigg[ \frac{g_2(1,\theta)}{g_3(1, \theta)^
{2/3}}\Bigg] \frac{1}{N ^{1/3}} \Bigg\}.  \label{Tc}
\end{eqnarray}
For $\theta =0$, the Dunkl-BEC temperature converts to the traditional
critical temperature form, $T_{c}^{B}$, given in \cite{gro1}. 
\begin{eqnarray}
T_c^B \simeq \frac{\hbar \Omega}{k_B} \bigg[\frac{N}{g_3(1)}\bigg]^ {1/3} %
\Bigg\{1-\frac{\gamma}{3} \Bigg[ \frac{g_2(1)}{g_3(1)^ {2/3}}\Bigg] \frac{1}{%
N ^{1/3}} \Bigg\}.  \label{Tb}
\end{eqnarray}
Then, we match Eqs. (\ref{Tc}) and (\ref{Tb}) to construct a relationship
between the latter and the conventional temperatures. We find the ratio 
\begin{equation}
\frac{T_{c}^{D}}{T_{c}^{B}}=\left[ \frac{\zeta \left( 3\right) }{%
g_{3}(1,\theta )}\right] ^{\frac{1}{3}}\frac{1-\frac{\gamma }{3N^{1/3}}\frac{%
g_{2}(1,\theta )}{g_{3}(1,\theta )^{2/3}}}{1-\frac{\gamma }{18N^{1/3}}\frac{%
\pi ^{2}}{\zeta \left( 3\right) ^{2/3}}},
\end{equation}
where $\zeta \left( n \right)$ is the Riemann-Zeta function. We note that
the second term in Eq. (\ref{Tc}) can be neglected if the particle number of
the ensemble is sufficiently large, $N\rightarrow \infty $. In this case,
the Dunkl-BEC temperature can be approximated by 
$T_{0}^{D}$%
\begin{equation}
T_{0}^{D}=\frac{\hbar \Omega }{k_B}\left[ \frac{N}{g_{3}(1,\theta )}\right]
^{\frac{1}{3}},  \label{TDC}
\end{equation}
which reduces to the ordinary case for $\theta=0$ \cite{gro1} 
\begin{equation}
T_{0}^{B}=\frac{\hbar \Omega }{k_B}\left[ \frac{N}{\zeta \left( 3\right) }%
\right] ^{\frac{1}{3}}.  \label{TBC}
\end{equation}
By comparing Eq. (\ref{TDC}) with Eq. (\ref{TBC}), we get the condensation
temperature ratio 
\begin{equation}
\frac{T_{0}^{D}}{T_{0}^{B}}=\left[ \frac{\zeta \left( 3\right) }{%
g_{3}(1,\theta )}\right] ^{\frac{1}{3}}.
\end{equation}
We display the change in the condensation temperature ratio versus the
Wigner parameter in Fig. \ref{fig3}. 
We see that this ratio increases with the increasing Wigner parameter value.
For the negative Wigner parameter values, this ratio is smaller than one. We
see that the ratio saturates at $1.794$ at large Wigner values.

\begin{figure}[htbp]
\centering
\includegraphics[scale=0.7]{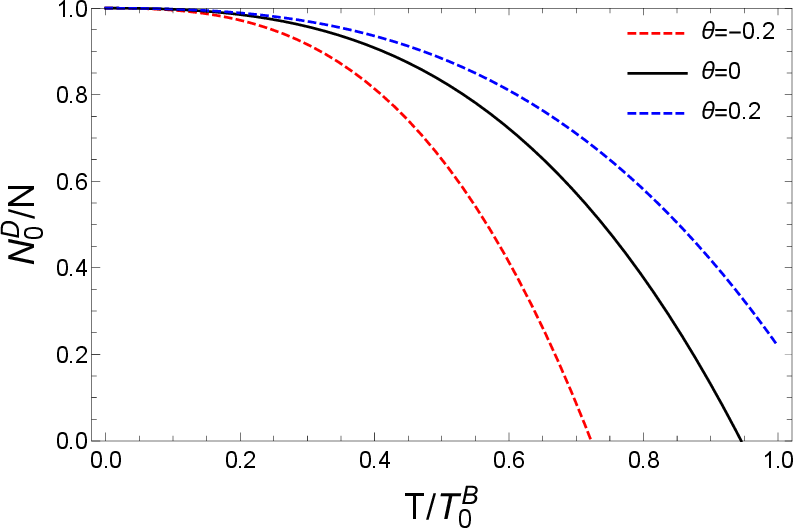}
\caption{The population of the Dunkl ground state ratio versus normalized
temperature for different Wigner parameters.}
\label{fig4}
\end{figure}

Then, by using Eq. \eqref{TBC} in Eq. \eqref{s}, we obtain the rate of Dunkl
ground state population in terms of normalized temperature as follows: 
\begin{equation}
\frac{N_{0}^{D}}{N}=1-\frac{g_{3}(1,\theta )}{\zeta \left( 3\right) }\left( 
\frac{T}{T_{0}^{B}}\right) ^{3}-\gamma \frac{g_{2}(1,\theta )}{\zeta
^{2/3}\left( 3\right) }\frac{1}{N^{1/3}} \left( \frac{T}{T_{0}^{B}}\right)
^{2}.
\end{equation}%
In Fig. \ref{fig4}, we plot this ratio versus the condensation temperature
ratio for an ensemble with two thousand particles. 

We see that for positive Wigner parameters, the ground state population of
the standard formalism is always smaller than the Dunkl formalism. In other
words, for $\theta >0$ the Dunkl-Bosonic system is more apt to undergo a
condensation than the standard-Bosonic system. In contrast, for negative
values of Wigner parameters, the ground state population in the Dunkl
formalism is smaller than the standard formalism.

\section{Thermodynamics of the system}

At this point, we can employ the internal energy, $U$ to derive the heat
capacity function. To obtain the internal energy we substitute the sum with
the weighted integral. In this case, the Dunkl internal energy 
\begin{equation}
U^D=2\int_{0}^{\infty }\frac{E\rho \left( E\right) dE}{e^{2\beta E}z^{-2}-1}%
+(1+2\theta )\int_{0}^{\infty }\frac{E\rho \left( E\right) dE}{e^{\beta
(1+2\theta )E}z^{-(1+2\theta )}+1}.  \label{u}
\end{equation}
yields the following result after the substituting Eq. (\ref{R}) into Eq. (%
\ref{u}): 
\begin{eqnarray}
\frac{U^{D}}{\hbar \Omega}&=& 3\left( \frac{k_B T}{\hbar \Omega }\right)
^{4}g_{4}(z,\theta )+2\gamma \left( \frac{k_B T}{\hbar \Omega }\right)
^{3}g_{3}(z,\theta ).
\end{eqnarray}

\begin{figure}[htbp]
\centering
\includegraphics[scale=0.7]{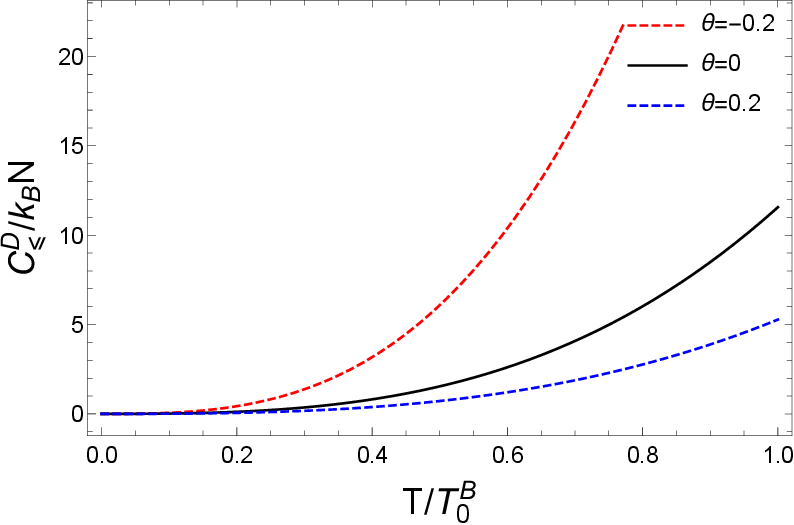}
\caption{Heat capacity function versus $\frac{T}{T_{0}^{B}}$ of a very large
number of bosons in the Dunkl formalism.}
\label{fig5}
\end{figure}

In order to compute the heat capacity, we have to make a distinction between
the two regimes. Below $T_{c}^{D}$, we may safely set $z=1$, however, for $%
T>T_{c}^{D}$ , we cannot since $z$ is a complicated function of $T$. By
using the well-known relation, $C=\partial U/\partial T$, we derive a
general expression for the reduced Dunkl heat capacity for $T<T_{c}^{D}$ 
\begin{equation}
\frac{C_{\leq }^{D}}{Nk_B}=12\frac{g_{4}(1,\theta )}{\zeta (3)}\left( \frac{T%
}{T_{0}^{B}}\right) ^{3}+\frac{6\gamma }{N^{1/3}}\frac{g_{3}(1,\theta )}{%
\zeta ^{2/3}(3)}\left( \frac{T}{T_{0}^{B}}\right) ^{2}\text{ },
\end{equation}
and for $T>T_{c}^{D}$ 
\begin{align}
\frac{C_{>}^{D}}{Nk_B}& =\frac{3}{2}\frac{g_{4}(z,\theta )}{\zeta \left(
3\right) }\left( \frac{T}{T_{0}^{B}}\right) ^{3}+\frac{3\gamma }{2N^{1/3}}%
\frac{g_{3}(z,\theta )}{\zeta ^{2/3}\left( 3\right) }\left( \frac{T}{%
T_{0}^{B}}\right) ^{2}  \notag \\
& +\left[ \frac{3}{4}\left( \frac{T}{T_{0}^{B}}\right) ^{4}\frac{%
g_{3}(z,\theta )}{\zeta \left( 3\right) }+\frac{\gamma }{N^{1/3}}\frac{%
g_{2}(z,\theta )}{\zeta ^{2/3}\left( 3\right) }\left( \frac{T}{T_{0}^{B}}%
\right) ^{3}\right] \frac{T_{0}^{B}}{z}\frac{dz}{dT}.
\end{align}
Here, the quantity $\frac{1}{z}\frac{dz}{dT}$ can be calculated by using the
fact that the total particle number is a constant. Considering $\frac{dN}{dT}%
=0$, we find 
\begin{equation}
\frac{T_{0}^{B}}{z}\frac{dz}{dT}=-3\frac{T_{0}^{B}}{T}.\frac{g_{3}(z,\theta )%
}{g_{2}(z,\theta )}\frac{1+\frac{2\gamma }{3}\frac{\zeta
^{1/3}\left(3\right) }{N^{1/3}}\frac{g_{2}(z,\theta )}{g_{3}(z,\theta )}%
\frac{T_{0}^{B}}{T}}{1+\gamma \frac{\zeta ^{1/3}\left( 3\right) }{N^{1/3}}%
\frac{g_{1}(z,\theta )}{g_{2}(z,\theta )}\frac{T_{0}^{B}}{T}},
\end{equation}
so that, the Dunkl-specific heat capacity reads 
\begin{align}
\frac{C_{>}^{D}}{Nk_B}& =\frac{3}{2}\frac{g_{4}(z,\theta )}{\zeta \left(
3\right) }\left( \frac{T}{T_{0}^{B}}\right) ^{3}+\frac{3\gamma }{2N^{1/3}}%
\frac{g_{3}(z,\theta )}{\zeta ^{2/3}\left( 3\right) }\left( \frac{T}{%
T_{0}^{B}}\right) ^{2} \\
& -3\frac{g_{3}(z,\theta )}{g_{2}(z,\theta )}\left[ \frac{3}{4}\left( \frac{T%
}{T_{0}^{B}}\right) ^{3}\frac{g_{3}(z,\theta )}{\zeta \left( 3\right) }+%
\frac{\gamma }{N^{1/3}}\frac{g_{2}(z,\theta )}{\zeta ^{2/3}\left( 3\right) }%
\left( \frac{T}{T_{0}^{B}}\right) ^{2}\right]\frac{ \frac{T}{T_{0}^{B}}+%
\frac{2\gamma }{3}\frac{\zeta ^{1/3}\left(3\right) }{N^{1/3}}\frac{%
g_{2}(z,\theta )}{g_{3}(z,\theta )}}{ \frac{T}{T_{0}^{B}}+\gamma \frac{\zeta
^{1/3}\left( 3\right) }{N^{1/3}}\frac{g_{1}(z,\theta )}{g_{2}(z,\theta )}}. 
\notag
\end{align}
Finally, in Fig.\ref{fig5} we plot the Dunkl heat capacity $C_{\leq }^{D}/KN$
versus the temperature $\frac{T}{T_{0}^{B}}$ for different values of Wigner
parameters. 

We observe that the heat capacity is a monotonically increasing function of $%
\frac{T}{T_{0}^{B}}$. Therefore, for a fixed value of $\theta $ and $N$, the
maximal value of $C_{\leq }^{D}/k_B N$ emerges around $T\simeq T_{0}^{B}$.
We observe that when considering negative Wigner parameters, the Dunkl heat
capacity surpasses the standard heat capacity. Similarly, in the scenario of
positive Wigner parameters, the standard formalism exhibits higher values
compared to the Dunkl formalism

\section{Conclusions}

We consider an ideal Bose gas trapped by a three-dimensional anisotropic
harmonic oscillator potential in the Dunkl formalism. We derive analytic
expressions of the Dunkl-BEC temperature and the Dunkl ground state
population and examine the dependence of these quantities on the Wigner
parameter. The internal energy and the heat capacity are also computed
analytically. The results are shown to nicely generalize known limiting
expressions in the non-deformed and/or untrapped cases

\section*{Acknowledgments}

This work is supp orted by the Ministry of Higher Education and Scientific
Research, Algeria under the co de: PRFU:B00L02UN020120220002. B. C. L\"{u}tf%
\"{u}o\u{g}lu is supported by the Internal Project [2023/2211], of Excellent
Research of the Faculty of Science of Hradec Kr\'{a}lov\'{e} University. 

\section*{Data Availability Statements}

The authors declare that the data supporting the findings of this study are
available within the article.

\end{document}